\documentclass[preprint]{elsarticle}

\usepackage{graphicx}
\usepackage{dcolumn}
\usepackage{bm}
\usepackage{multirow}
\usepackage{ulem}
\usepackage{soul}
\usepackage{subfig}
\usepackage{array}
\usepackage{lineno}
\usepackage{longtable}
\usepackage{rotating}
\newcolumntype{P}[1]{>{\centering\arraybackslash}p{#1}}
\begin{document}

\title{Laboratory investigations of Lunar ice imaging in permanently shadowed regions using reflected starlight}
\author[YRK]{Paul J. Godin}
\author[YRK]{Jacob L. Kloos}
\author[YRK]{Alex Seguin}
\author[YRK]{John E. Moores}
\address[YRK]{Department of Earth and Space Science and Engineering, York University, 4700 Keele St., Toronto, ON, M1J 1P3, Canada}

\begin{abstract}

A proof of concept for a frost detection imager using reflected starlight is presented; the limitations of this technique are explored experimentally. An ice-covered lunar surface is simulated inside a vacuum chamber, which is then illuminated with a lamp containing UV and visible output to simulate the wavelengths of the background starfield. The simulated lunar surface is imaged with a camera utilizing a UV and visible filter pairing. At Lyman-$\alpha$ wavelengths, ice has low reflectivity, and on average appears darker than the regolith in the UV image. In visible wavelengths, this behaviour is reversed, with ice appearing brighter than regolith. 
\\ \indent
UV/VIS image ratioing is subsequently performed in order to discern frost from the lunar regolith simulant in order to demonstrate the capability of this technology for locating the presence of ice on the lunar surface. When the two images are ratioed, the signal to noise ratio to distinguish ice from regolith improves by 36\%. In cases where the presence of shadows and specular reflection make distinguishing ice from regolith in either a single UV or visible image difficult, ratioing the images makes the distinction clear. 

\end{abstract}
\date{\today}
\maketitle

\section{Introduction}
\label{sec:Intro}

Permanently Shadowed Regions (PSRs) on the moon can function as traps for water, since temperatures remain low enough to preserve volatiles for extended periods in these locations \cite{Paige, Paige2}. Ice can be delivered directly by impacts into the PSRs, collected from the solar wind \cite{Sunshine, Dyar, McCord}, or transported by ballistic migration from the equator \cite{Schorghofer, Moores} and pumped into the subsurface \cite{Schorghofer2}.  Water stored in the PSRs has high scientific and exploration value. For science, this reservoir may preserve a record of bulk water delivered to planetary surfaces by cometary impacts. For exploration, this water ice offers a potential readily available source of hydrogen and oxygen that can be used to sustain exploration activities; the water itself can be used by astronauts directly as potable water, the oxygen can be used to manufacture breathable air and the hydrogen can be used as a feedstock to produce fuel for operations and the eventual return to the Earth.
\\ \indent
Spacecraft have observed a suppression of epithermal neutron flux, which can indicate the increased presence of hydrogen, suggesting significant deposits of water ice at these PSRs \cite{Mitrofanov, Mitro2, Feldman}. The LCROSS impact vaporization experiment also detected significant water in Cabeus crater \cite{Colaprete}; and measurements by the Moon Mineralogy Mapper spectrometer on board the Chandrayaan-1 spacecraft have observed water ice using reflected IR absorption features \cite{Bandfield, Li}. The Lunar Orbiter Laser Altimeter (LOLA) aboard Lunar Reconnaissance Orbiter (LRO) has used surface albedo at 1064 nm to infer the presence of ice deposits inside PSRs \cite{Qiao, Rubanenko, Nash}. Characterizing ice deposits on the lunar surface is considered such a high priority that both NASA and ESA are planning new satellite mission for detect lunar frost using active remote sensing in the IR; Lunar Flashlight and Lunar Volatile and Mineralogy Mapping Orbiter, respectively.
\\ \indent
Another solution is to use the existing illumination provided by starlight. All stars have significant output at Lyman-$\alpha$ wavelengths in the Vacuum Ultraviolet (VUV) due to the hydrogen in their photospheres. There is also a significant glow produced by interplanetary hydrogen along the interstellar breeze. The observation of frosts at Haworth crater by LRO-LAMP \cite{Hayne,Gladstone} relies on this principle and demonstrates that frost may be detected using natural Lyman-$\alpha$ illumination by a detector sensitive at 121.6 nm. Reflectance data from the Apollo samples measured by Hapke \textit{et al.} \cite{Hapke} show that lunar regolith has an albedo of 3-5\% in the deep UV. Water ice however has an albedo of 1-3\% in the deep UV \cite{Gladstone, WB, Jake}. LAMP builds up reflectance data from multiple fly overs of PSRs to search for this small, but detectable change in surface albedo to distinguish ice from regolith. A similar method can be applied using visible wavelengths: studies have show that there is significant visible light emission in the PSRs from the background starfield \cite{Jake} and Earthshine is another source of illumination in the PSRs \cite{Jake, Jake2, Glenar}. Over visible wavelengths, lunar regolith averages an albedo of $\sim$10\% \cite{Hapke}, while water ice has an albedo of 80-90\% in the visible \cite{Hapke,WB, Jake}. The upcoming Korea Pathfinder Lunar Orbiter ShadowCam plans to survey PSRs using reflected light.
\\ \indent
An improved method for detecting ice in PSRs is possible by considering both visible and UV light at the same time. A ratio of UV and visible images would increase the relative reflectance difference, resulting in an order of magnitude difference in brightness between ice and regolith, an would provide a clearer difference between ice and regolith than simply using UV or visible imaging alone. This paper reports laboratory simulations of lunar frost detection using ratios of reflected UV and visible light as a proof of concept for a rover mounted camera for future lunar frost detection missions. 

\section{Experimental Methods}
\label{sec:Meth}
\begin{figure}
\includegraphics[width=1.0\textwidth]{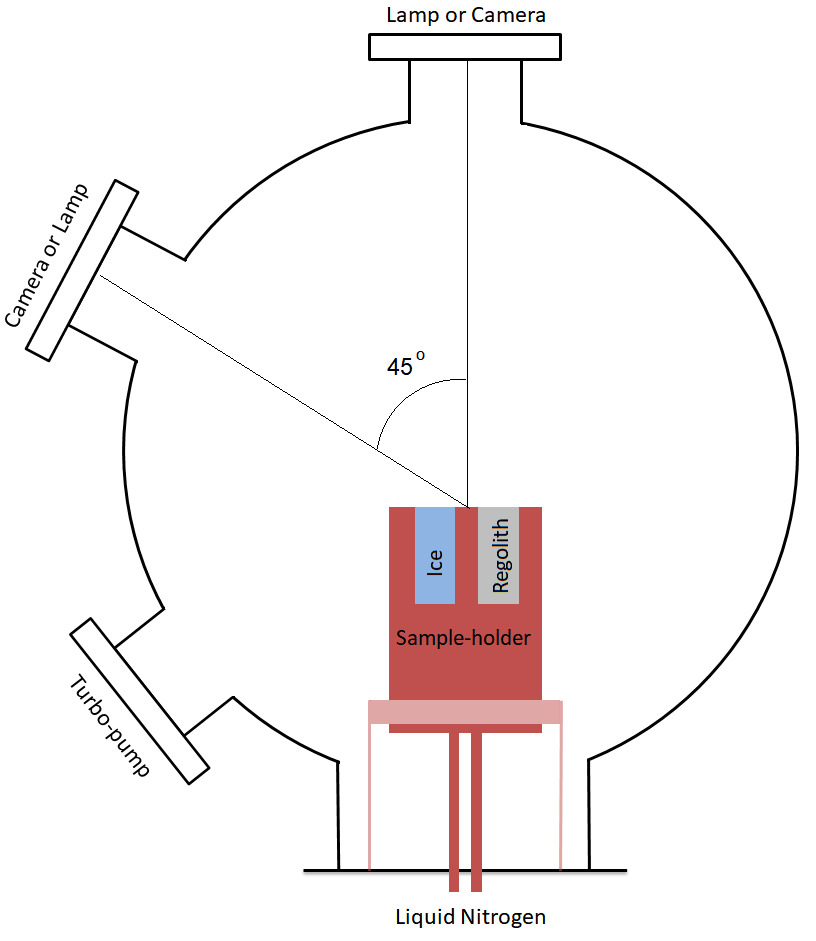}
\caption{Schematic diagram of the experimental set-up. A copper sample holder is placed such that the top of the holder is at the centre of the chamber. The holder is in physical contact with a liquid nitrogen heat exchanger. The camera and lamp are installed on two flanges and are aimed at the sample holder; the angle made between the lamp, sample, and camera is 45$^{o}$. A third flange leads to the turbo-pump.}
\label{fig:SU}
\end{figure}

The lunar surface was simulated inside of a 12" diameter cryo-vacuum chamber. Liquid nitrogen was pumped through a heat exchanger to achieve temperatures of 95 K at a copper sample holder in the centre of the chamber. Pressure control was provided by a turbomolucular pump, which was able to achieve pressures on the order of 10$^{-5}$ Torr. JSC-1A lunar regolith simulant was used to simulate lunar regolith and deionized water was frozen to create ice. Ice and regolith were placed inside the sample holder. Two sample locations were filled with regolith, the other two sample locations were filled with crushed ice. Mixtures of ice and regolith were not investigated because preliminary testing found that ice would sublimate from mixtures, thus changing the value of the mixing ratio. The pressure in the chamber was not reduced until the temperature was far below freezing. This was done to prevent the ice from sublimating once the pressure is lowered. Temperature and pressure take approximately an hour to stabilize to the desired levels. Once the temperature and pressure of the system were stabilized, UV and visible images were taken one immediately after the other. A schematic diagram of the experimental set-up is shown in Figure \ref{fig:SU}. To test the effects of different light distributions, the camera and lamp positions were interchanged to obtain two sets of images of the same sample under different viewing geometries. 

\begin{figure}
\includegraphics[width=1.0\textwidth]{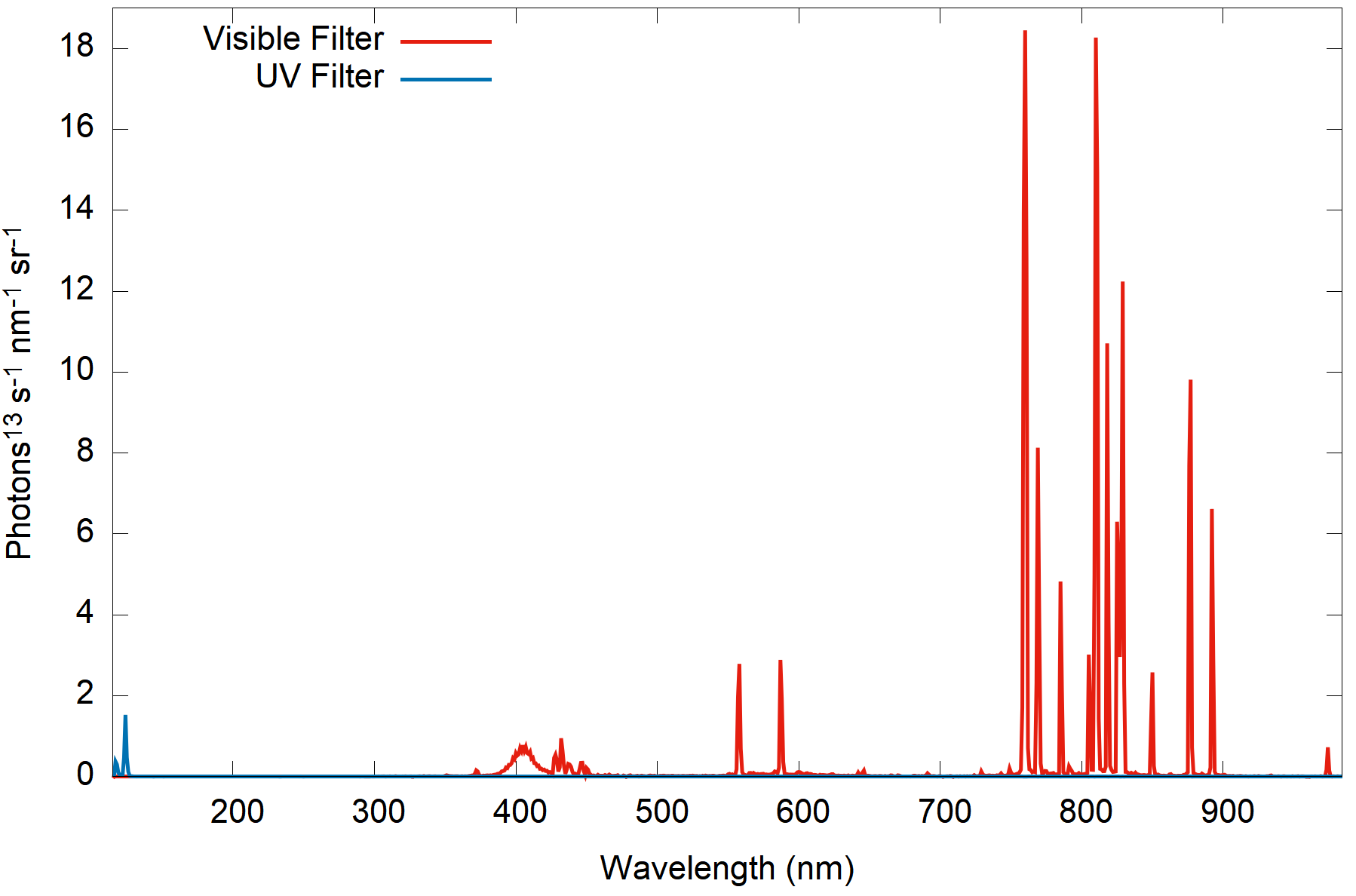}
\caption{The combined effect of lamp output, quantum efficiency, and filters on the spectrum observable by the camera.}
\label{fig:Cam_spec}
\end{figure}

\indent
A Resonance Ltd. Krypton line source VUV lamp was used to simulate starlight. This light source was chosen because it outputs significantly in the deep UV (115-130 nm) and broadband visible light (400-1000 nm). Images were taken with a Resonance Ltd. VUVCam1-121 Ultraviolet Camera, which is capable of imaging at both wavelength regimes produced by the lightsource since the quantum efficiency of the camera is $\sim30\%$ in the deep UV and 10-80\% over the broadband visible range. In order to ratio images at two different spectral regions, a band pass filter was placed in front of the camera. Two different filters were used, an MgF$_{2}$ filter for Lyman-$\alpha$ and a N-BK7 glass lens for visible. A lens was used as because the camera was optimized for UV operation, so the built in lens was insufficient to focus visible wavelengths. The combined effect of lamp output, quantum efficiency, and filters on the spectrum observable by the camera shown in Figure \ref{fig:Cam_spec}. Sources of uncertainty include MgF$_{2}$ filter transmission ($\pm1\%$), N-BK7 glass filter transmission ($\pm0.1\%$), lamp fluctuations ($\pm10\%$), and camera QE ($\pm5\%$). In order to reduce uncertainty, 100 dark and image pairs were taken in both UV and visible. The raw images were individually dark subtracted before the 100 images were averaged together.

\section{Data Analysis}
\label{sec:DA}

Image preparation before ratioing was needed to account for the optical focus of the images at different wavelengths. The different wavelengths between the two images result in a different area being imaged onto the CMOS sensor, creating the appearance that the visible image is zoomed-in compared to the UV image. This can been seen by comparing the amount of the sample holder contained in the raw images in Figure \ref{fig:images}. To account for this optical effect, the UV image is cropped to match the visual scene of the image seen in the visible, then the pixel values of the visible image are averaged to shrink the visible image pixel count to match the number of pixels in the cropped UV image. This is a source of uncertainty in this technique, if an optical path could be designed that would produce an identical image in both UV and visible, uncertainty be further reduced. A 3x3 pixel filter is passed over the images to further reduce noise in the images.
\\ \indent
Due to the low signal intensity from the camera at Lyman-$\alpha$, UV images were taken at a gain of 29 dB, and an exposure time of 3800 ms. By contrast, due to high signal levels in the visible, images were taken at a gain of 0 dB and an exposure time of 35 ms. Multiplying the spectra in Figure \ref{fig:Cam_spec} and the respective gains and exposures, and integrating over the spectral region, results in the detectable UV signal being 47 times stronger than the visible signal, assuming an equally reflective surface. This correction factor was applied to the UV images, so that when the UV and visible images are ratioed, they are on equal footing, thus any differences in brightness are due to surface reflectivity. 

\section{Results}
\label{sec:Results}

\begin{figure}
\begin{center}
\subfloat[Raw visible top-down image.]{\label{fig:VisT}\includegraphics[width=0.45\textwidth]{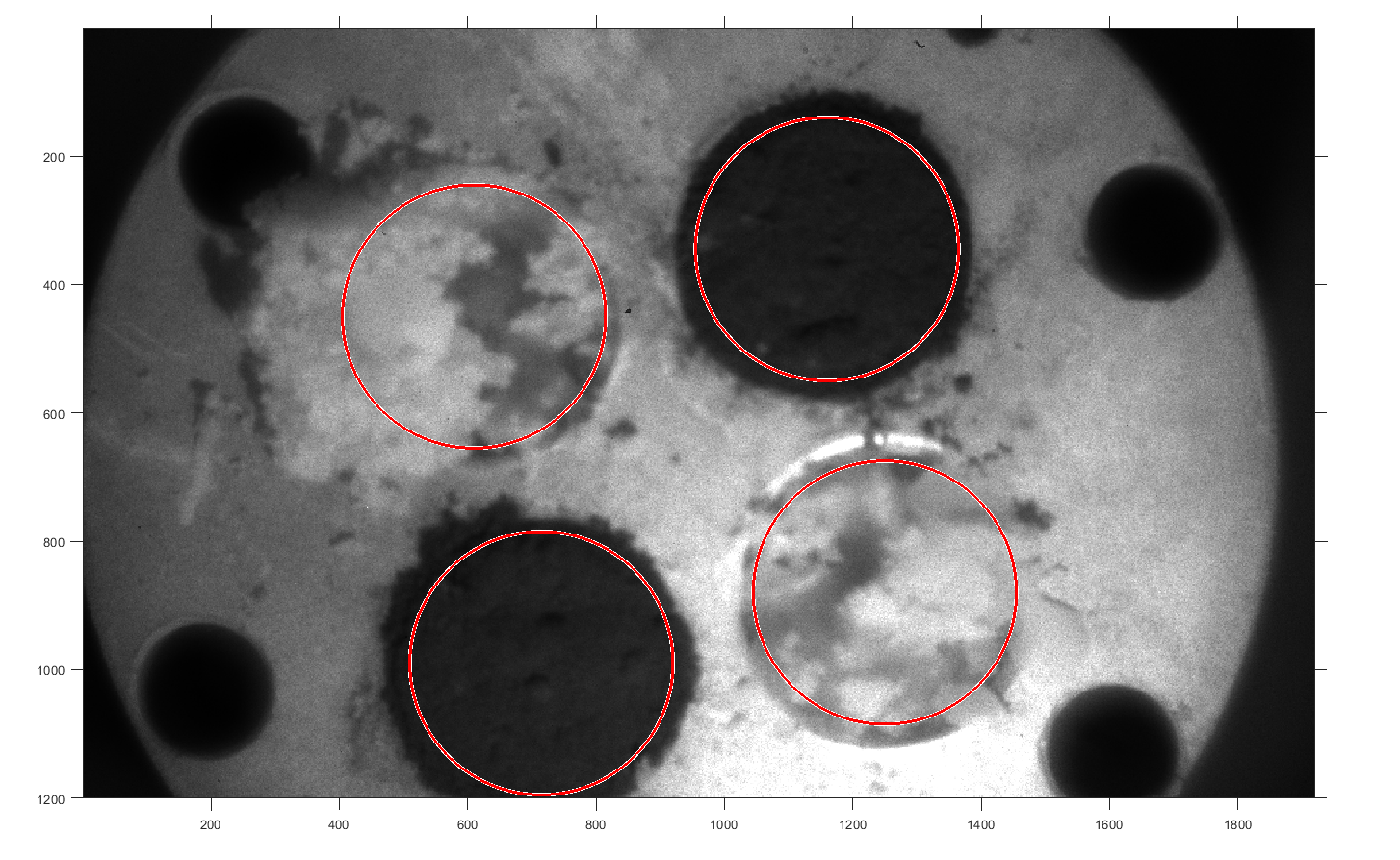}} 
\subfloat[Raw visible sideways image.]{\label{fig:VisS}\includegraphics[width=0.45\textwidth]{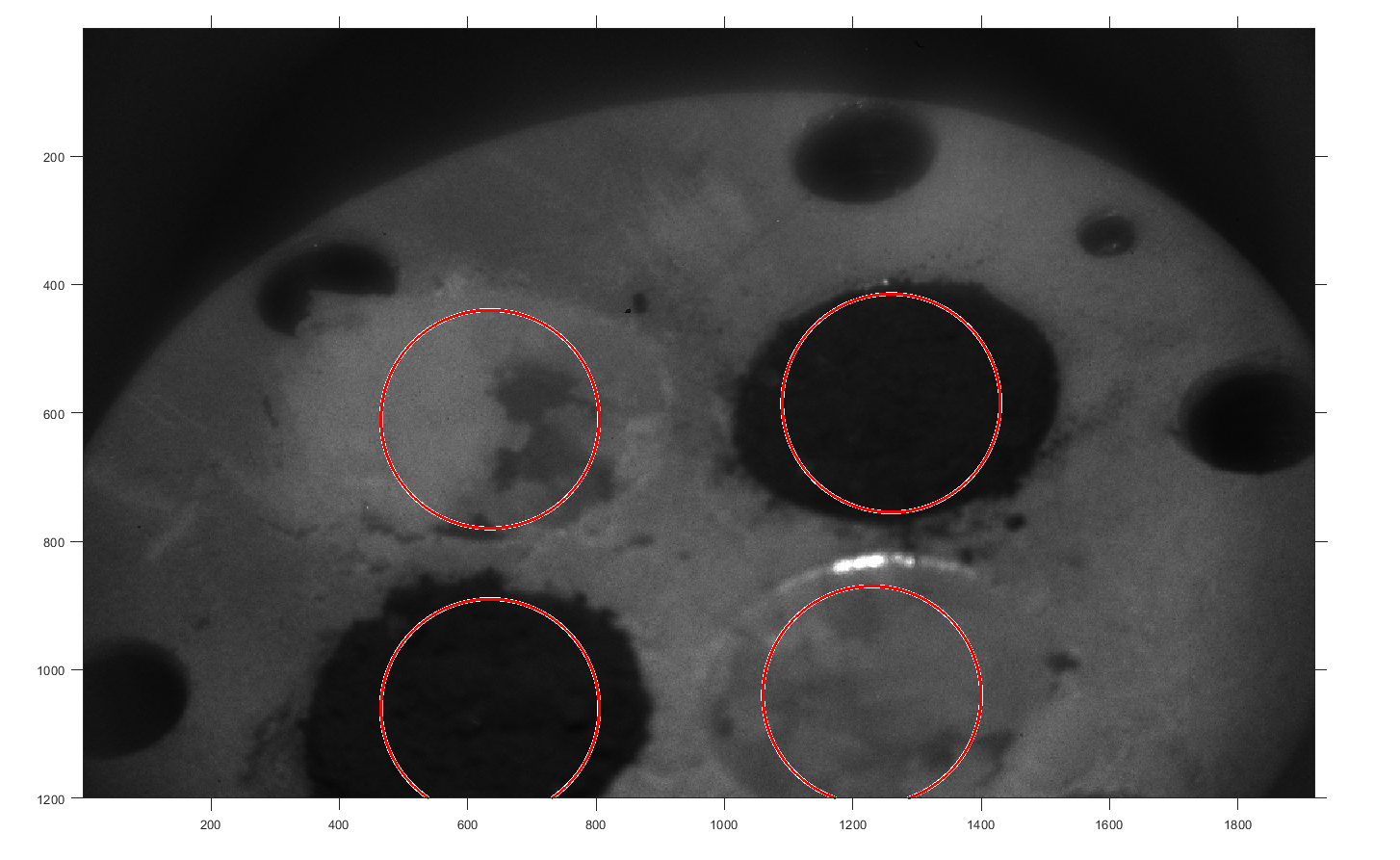}} 
\\
\subfloat[Raw UV top-down image.]{\label{fig:UVT}\includegraphics[width=0.45\textwidth]{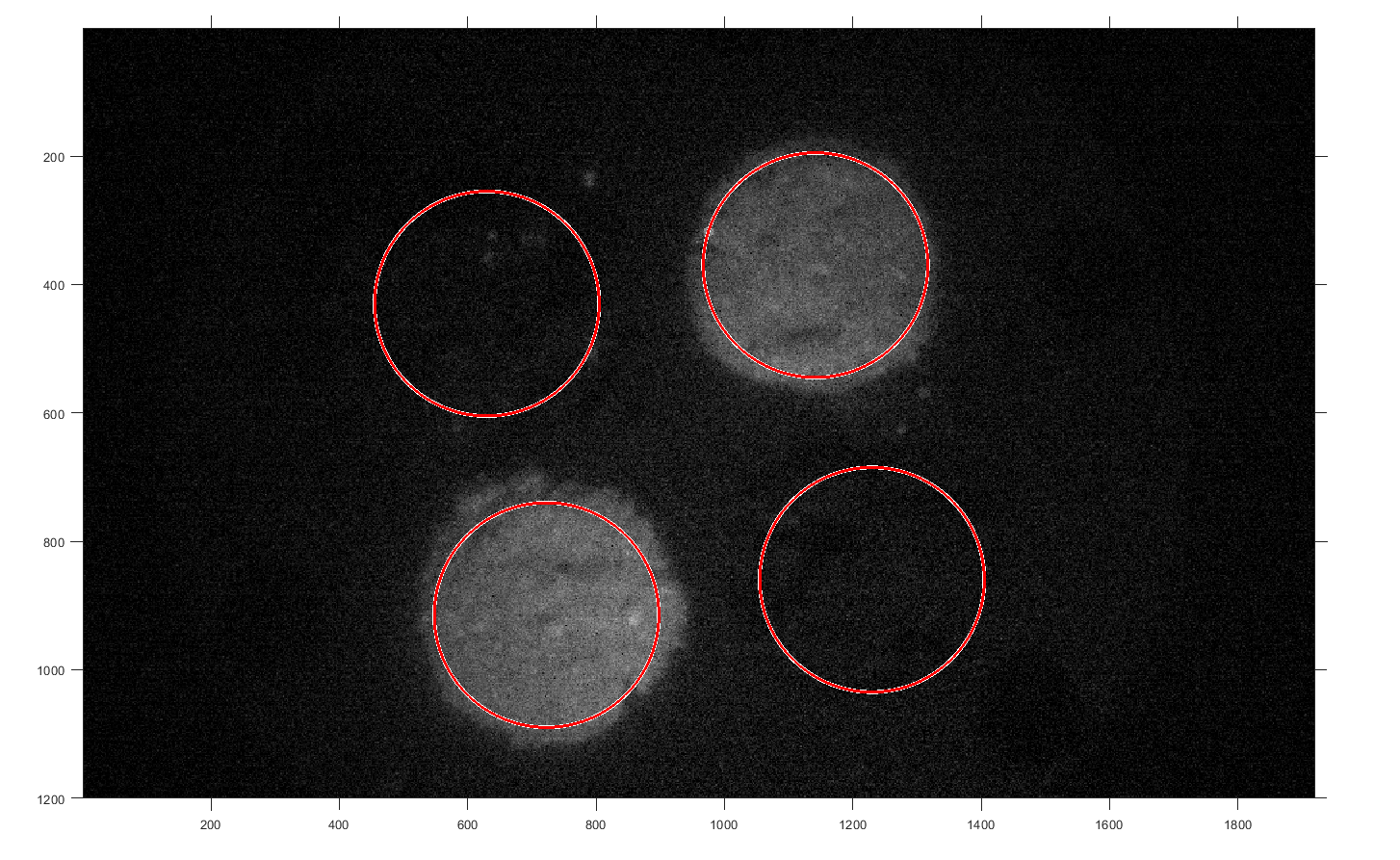}} 
\subfloat[Raw UV sideways image.]{\label{fig:UVS}\includegraphics[width=0.45\textwidth]{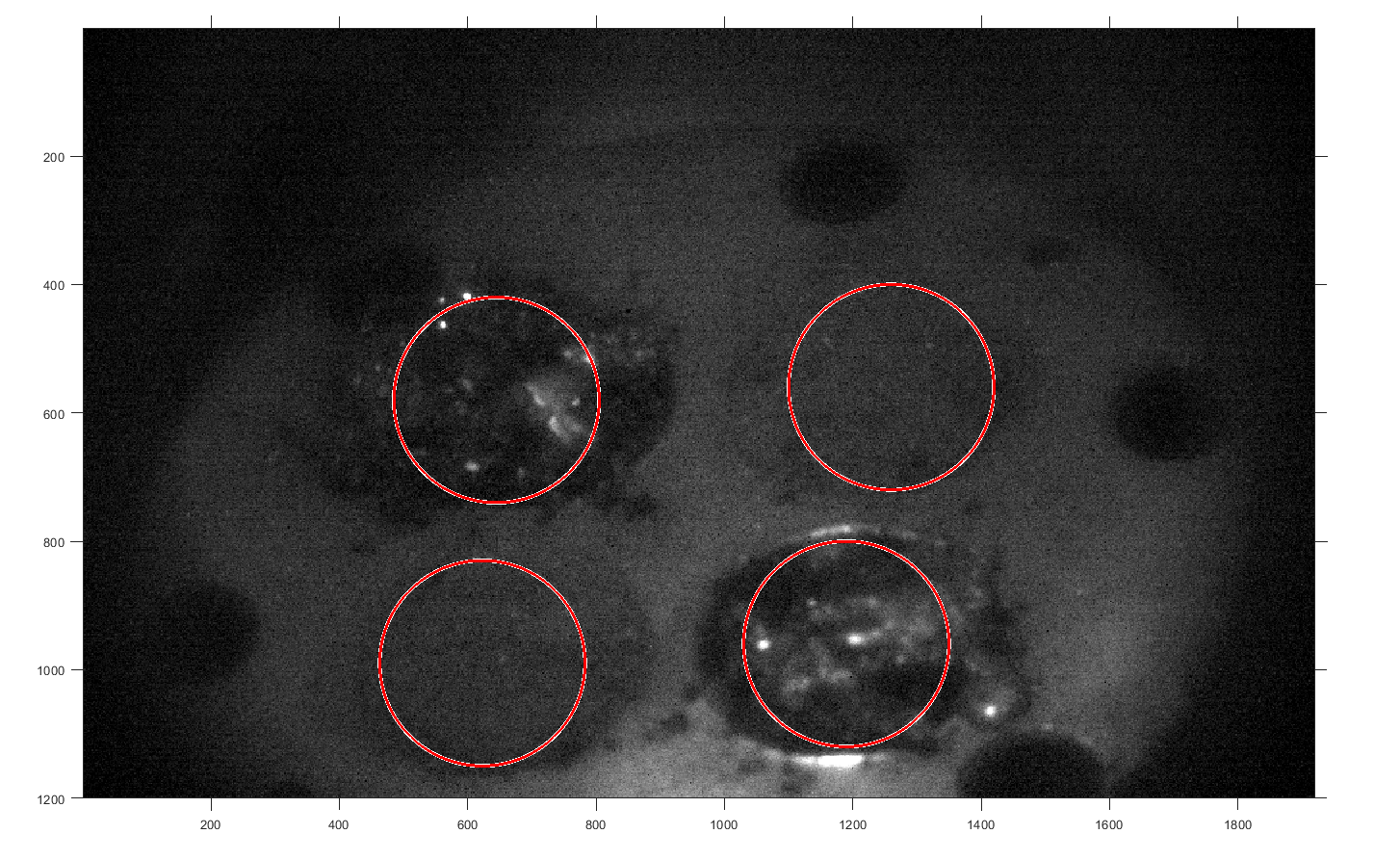}} 
\\
\subfloat[Ratio of scaled top-down images.]{\label{fig:RT}\includegraphics[width=0.45\textwidth]{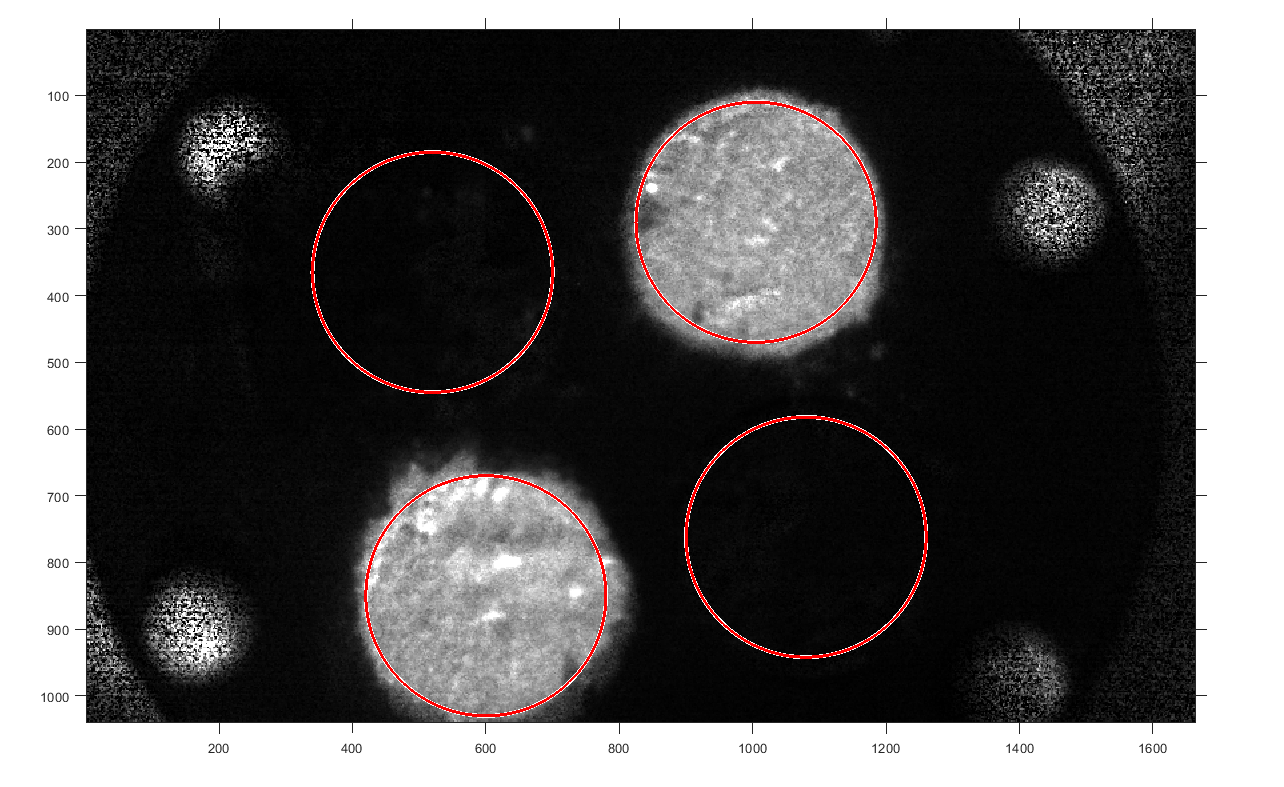}} 
\subfloat[Ratio of scaled  sideways images.]{\label{fig:RS}\includegraphics[width=0.45\textwidth]{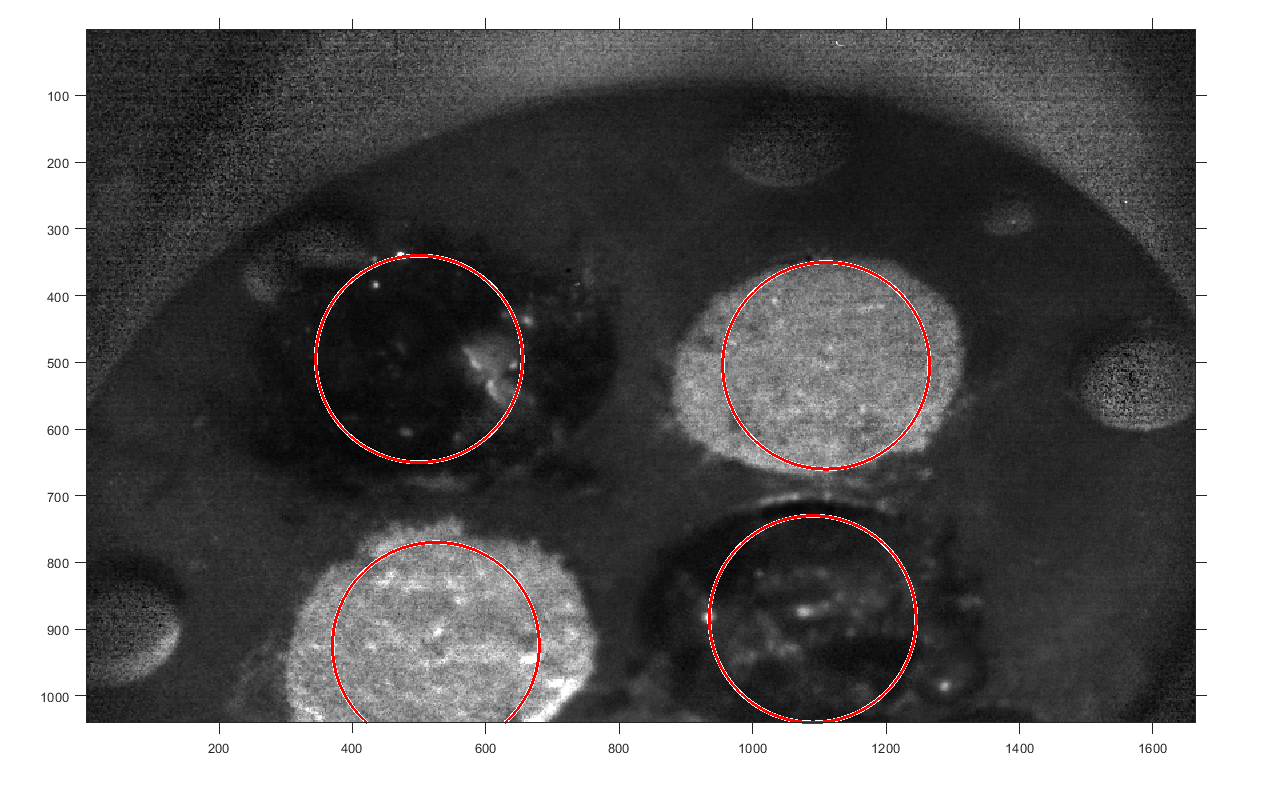}}
\\
\caption{Images of regolith and ice samples. Top left and bottom right locations in each image are ice samples. Top right and bottom left locations in each image are regolith. Red circles identify areas used to calculate average sample reflectance.}
\label{fig:images}
\end{center}
\end{figure}

\begin{figure}
\begin{center}
\subfloat[Top-down UV Only.]{\label{fig:UVTH}\includegraphics[width=0.45\textwidth]{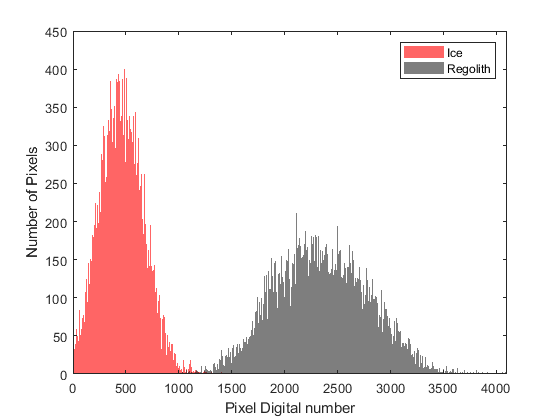}} 
\subfloat[Sideways UV Only.]{\label{fig:UVSH}\includegraphics[width=0.45\textwidth]{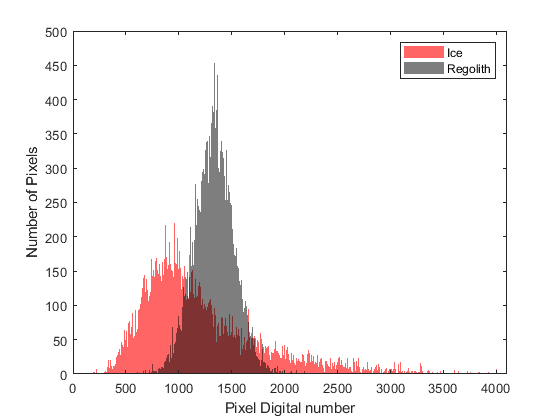}} 
\\
\subfloat[Top-down Visible Only.]{\label{fig:VisTH}\includegraphics[width=0.45\textwidth]{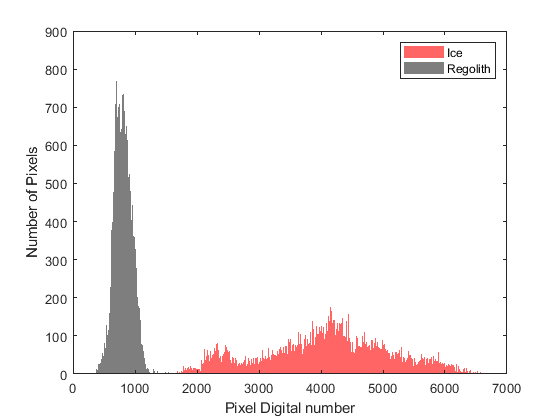}} 
\subfloat[Sideways Visible Only.]{\label{fig:VisSH}\includegraphics[width=0.45\textwidth]{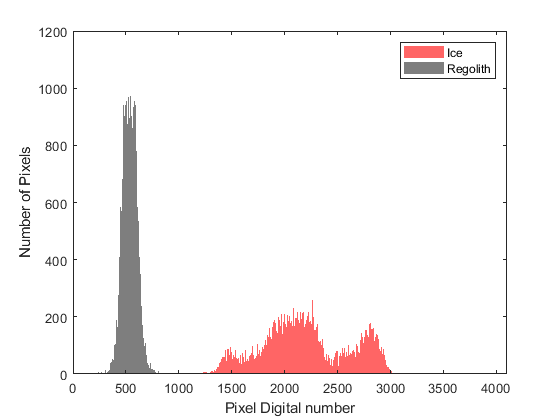}} 
\\
\subfloat[Top-down UV/Vis Ratio.]{\label{fig:RTH}\includegraphics[width=0.45\textwidth]{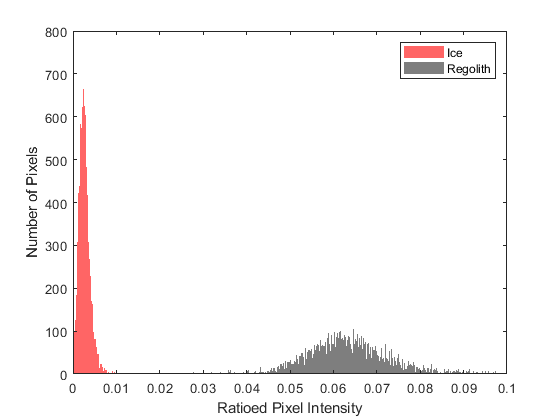}} 
\subfloat[Sideways UV/Vis Ratio.]{\label{fig:RSH}\includegraphics[width=0.45\textwidth]{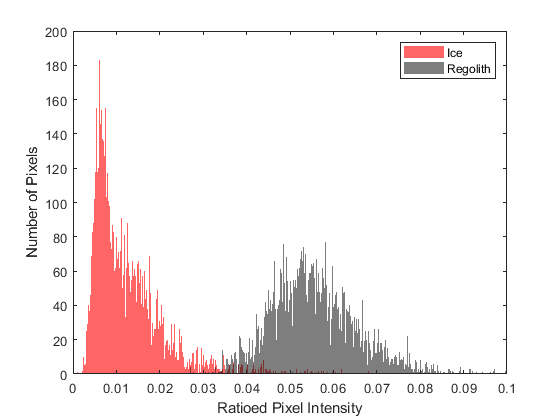}} 
\\
\caption{Number of pixels in sample area as a function of pixel intensity. A scaling factor of 47 has been applied to the ratioed images to account for differences in camera settings/light intensities.}
\label{fig:RR}
\end{center}
\end{figure}

Images before and after processing are shown in Figure \ref{fig:images}. Top left and bottom right locations in each image are ice samples. Top right and bottom left locations in each image are regolith. Significant differences in grain size can be seen in the ice samples, resulting in non-uniform appearance across the sample surface. The regolith samples have much less variability in grain size and distribution, yielding more consistent results than the ice samples. The amount of noise in the images was determined by looking at the variability of pixels in the far corners of the images. For the UV images, the standard deviation of corner pixels was $\sim127$ digital pixel numbers; while for the visible images it was $\sim67$. The additional variability seen in Figure \ref{fig:RR} and be attributed to variation of the sample surfaces.

To confirm the camera's capabilities, the relative brightness of ice and regolith in an individual image was compared. This was performed by selecting areas of the images containing ice or regolith samples (designated by the circles in Figure \ref{fig:images}), and calculating the median pixel value and uncertainty at two standard deviations. The ratio of ice to regolith albedos in this work can be compared to published values. For regolith, the albedo is taken as an average of the Apollo samples measured by Hapke \textit{et al.} \cite{Hapke}. For ice, there are multiple modeled reflectance data sets available; there is the directional hemispheric reflectance work by Kloos \textit{et al.} \cite{Jake} and bi-directional reflectance modeling using optical constants from Warren and Brandt \cite{WB}. While there exist experimental measurements of UV albedo of ice \cite{Hapke}, there are potential issues with these measurements, such as red leak and contamination \cite{Goguen}. In the deep UV, the modeled work \cite{WB, Jake} both predict $\sim$0.01. In the visible, there is general agreement, with both experimental and modeled work predicting little variation with wavelength and albedos of $\sim$0.8. To validate the ice to regolith reflectance ratios seen in this work to the literature, a convolution of the ratio of ice and regolith albedos from the literature and the observable spectrum in this experiment (Figure \ref{fig:Cam_spec}) was preformed to yield a predicted brightness ratio. This convolution is given as: 

\begin{equation}
R = \frac{\int{\alpha_{R}\sigma_{f}}}{\int{\sigma_{f}}}
\label{eq:convolve}
\end{equation}
\noindent

where $R$ is the predicted brightness ratio of ice to regolith observable in this experiment, $\alpha_{R}$ is the ratio of ice albedo (averaged from \cite{WB} and \cite{Jake}) to regolith albedo (taken from \cite{Hapke}), and $\sigma_{f}$ is the lamp spectrum seen by the camera with the subscript denoting either the visible or UV filter. These results are summarized in Table \ref{tab:Ice/Reg}. Good agreement is seen for albedo ratios in the visible compared with the literature. In the UV however, there is significant variability in the experimental UV measurement, which appears to be caused by scattering off of non-uniform ice grain sizes and specular reflection. Despite this, the UV albedo ratios agree with the literature once averaged. 

\begin{table}
\begin{center}
\caption{\label{tab:Ice/Reg} Summary of albedo ratios for ice/regolith. Uncertainties are 2$\sigma$.}
\begin{tabular}{|P{1.8cm}|P{2.3cm}|P{2.3cm}|P{2.2cm}|P{1.8cm}|} \hline
\textbf{Image Set} & \textbf{Top-down} & \textbf{Sideways} & \textbf{Average} & \textbf{Literature*} \\ \hline
UV & $0.1922\pm0.0004$ & $0.816\pm0.003$ & 0.504$\pm$0.003 & 0.55$\pm$0.14 \\ 
Visible & $5.139\pm0.006$ & $4.030\pm0.004$ & 4.585$\pm$0.007 & 5.32$\pm$1.37 \\  
Ratio & $0.0385\pm0.0001$ & $0.1897\pm0.0008$ & 0.1141$\pm$0.0008 & 0.103$\pm$0.037 \\ \hline
\end{tabular}
\\
 **Average of ice albedos from \cite{WB} and \cite{Jake}, and regolith albedo from \cite{Hapke}.  
\end{center}
\end{table}

It should be noted that the albedos used by Kloos \textit{et al.} \cite{Jake} are directional hemispheric albedos, which is a ratio of the power of reflected light in the entire hemisphere to the power of light incident from a single direction. By the principal of reciprocity, this is the type of albedo that is relevant for a camera looking into a PSR given that starlight is diffuse and coming from all directions. However, the experimental results reported here are bi-directional, given that there is only a single illumination source. The difference between hemispheric and bi-directional could explain why most of the experimental ratios are smaller than the predicted ratios.
\\ \indent
Given that the experimental set-up was behaving as expected based on ice versus regolith albedo ratios, the next step was to look at ratios of UV to visible images. After ratioing the UV and visible images, the regolith appears significantly brighter, compared to the ice, as seen in Figures \ref{fig:RT} and \ref{fig:RS}. To better understand the uncertainties in this detection technique, histograms showing the distribution of pixel intensities for ice and regolith are shown in Figure \ref{fig:RR}. In general it can be seen that both ice and regolith nave normal distributions of pixel brightness, with ice appearing darker than regolith in the UV, while the reverse is true for the visible images. The one notable exception is in the sideways view, where in the UV ice has a long tail off of brighter pixels, which appears to be caused by specular reflection off of the ice; while in the visible, the ice has significant variability. This appears to be caused by variation in grain size producing shadows and reflections in the ice samples. Figure \ref{fig:RR} also shows clear improvement in the ability to distinguish ice from regolith using a ratio of UV to visible, over UV alone. 
\\ \indent
Finally, the signal to noise ratio (SNR) to distinguish ice from regolith is calculated. The SNR is defined as:

\begin{equation}
SNR = \frac{\left|I-R\right|}{\sqrt{\sigma_I^{2}+\sigma_R^{2}}}
\label{eq:SNR}
\end{equation}
\noindent
where $I$ is the median ice pixel value, $R$ is the median regolith pixel value, $\sigma_I$ is the 95\% confidence interval of ice pixels, and $\sigma_R$ is the 95\% confidence interval of regolith pixels. A summary of the SNR derived from the images are listed in Table \ref{tab:SNR}. In cases where ice and regolith are distinguishable in both UV and visible, such as the top-down images, ratioing the images improves the SNR by 36\%. In cases where ice and regolith are difficult to distinguish in either UV or visible, such as the sideways images, ratioing the images makes the distinction between ice and regolith clear, as shown in Figure \ref{fig:RSH}. Ultimately, ratioing UV to visible images shows significant improvement for ice detection over UV or visible alone.

\begin{table}
\begin{center}
\caption{\label{tab:SNR} Summary of SNR for ice to regolith detection.}
\begin{tabular}{|P{3.5cm}|P{1.5cm}|} \hline
\textbf{Data Set} & \textbf{SNR} \\ \hline
Sideways UV & 71.6 \\ 
Top-down UV & 840.6 \\ \hline
Sideways Visible & 841.8 \\ 
Top-down Visible & 827.6 \\ \hline 
Sideways Ratio & 635.7 \\ 
Top-down Ratio & 1133.3 \\ \hline
\end{tabular} 
\end{center}
\end{table}

The SNR values presented here are quite promising, however for a camera on the lunar surface, the actual SNR will depend on the amount of light and optics used. Kloos \textit{et al.} \cite{Jake} what the SNR of the camera used in this study would be using the light levels they simulate in their paper. They find that reasonable SNR can be achieved at short integration times ($<0.1$ seconds), for a single image. The SNR could be further improved with repeated measurements or larger aperture optics than the one used in this study. 

\section{Conclusions}
\label{sec:Conc}
Laboratory simulations of the lunar surface have demonstrated that distinguishing lunar ice from lunar regolith is possible using a ratio of UV and visible reflectance images, due to strong absorption of deep UV radiation by water ice. Using a ratio of UV to a visible image improves the SNR of a detection compared to using UV or visible alone. When the two images are ratioed, the signal to noise ratio to distinguish ice from regolith improves by 36\%. In cases where the presence of shadows and specular reflection make distinguishing ice from regolith in either a single UV or visible image difficult, ratioing the images makes the distinction clear. If such imaging optics are used from the perspective of a rover on the surface of the moon, a much higher resolution map of frost may be obtained, since the proximity to the ice is smaller than from orbit and longer exposures are possible if the rover is stationary during imaging.


\section*{References}

\begin{thebibliography}{99}

\bibitem{Paige}
 D. A. Paige, and 22 co-authors, The Lunar Reconnaissance Orbiter Diviner Lunar Radiometer Experiment, \textit{Space Sci Rev.}, doi: 10.10047/s11214-009-9529-2 (2009).

\bibitem{Paige2}
 D. A. Paige, and 26 co-authors, Diviner Lunar Radiometer Observations of Cold Traps in the Moon’s South Polar Region, \textit{Science}, \textbf{330}, p.479-482 doi: 10.1126/science.1187726 (2010).

\bibitem{Sunshine}
J. M. Sunshine, T. L. Farnham, L. M. Feaga, O. Groussin, F. Merlin, R. E. Milliken and M. F. A’Hearn, Temporal and Spatial Variability of Lunar Hydration As Observed by the Deep Impact Spacecraft. \textit{Science}, \textbf{326}, 5952, p. 565-568  (2009).

\bibitem{Dyar}
M. D. Dyar, C. A. Hibbitts and T. M. Orlando, Mechanisms for incorporation of hydrogen in and on terrestrial planetary surfaces. \textit{Icarus}, \textbf{208}, 1, p. 425-437  (2011).

\bibitem{McCord}
T. B. McCord, L. A. Taylor, J.-P. Combe, G. Kramer, C. M. Pieters, J. M. Sunshine, and R. N. Clark, Sources and physical processes responsible for OH/H2O in the lunar soil as revealed by the Moon Mineralogy Mapper (M3), \textit{J. Geophys. Res.}, \textbf{116}, E00G05 doi: 10.1029/2010JE003711 (2011).

\bibitem{Schorghofer}
N. Schorghofer, Migration calculations for water in the exosphere of the Moon: Dusk-dawn asymmetry, heterogeneous trapping and D/H fractionation, \textit{Geophys. Res. Lett.}, \textbf{41}, 14 pp. 4888-4893. Doi: 10.1002/2014GL060820 (2014).

\bibitem{Moores}
J. E. Moores, Lunar water migration in the interval between large impacts: Heterogeneous delivery to Permanently Shadowed Regions, fractionation, and diffusive barriers, \textit{J. Geophys. Res. Planets}, \textbf{121}, p.46–60, doi:10.1002/2015JE004929 (2016).

\bibitem{Schorghofer2}
N. Schorghofer and O. Aharonson, The Lunar Thermal Ice Pump, \textit{Ap. J.}, \textbf{788}, 2, p.169, doi:10.1088/0004-637X/788/2/169 (2014).

\bibitem{Mitrofanov}
I. Mitrofanov, and 28 co-authors, Hydrogen Mapping of the Lunar South Pole Using the LRO Neutron Detector Experiment LEND, \textit{Science}, \textbf{330}, 6003, p.483, doi: 10.1126/science.1185696 (2010).

\bibitem{Mitro2}
I. Mitrofanov, and 23 co-authors, Testing polar spots of water-rich permafrost on the Moon: LEND observations onboard LRO, \textit{J. Geophys Res.}, \textbf{117}, E00H27, doi: 10.1029/2011JE003956 (2012).

\bibitem{Feldman}
W. C. Feldman, S. Maurice, A. B. Binder, B. L. Barraclough, R. C. Elphic, and D. J. Lawrence Fluxes of Fast and Epithermal Neutrons from Lunar Prospector: Evidence for Water Ice at the Lunar Poles, \textit{Science}, \textbf{281}, 5382, p.1496-1500 (1998).

\bibitem{Colaprete}
A. Colaprete, and 16 co-authors, Detection of Water in the LCROSS Ejecta Plume, \textit{Science}, \textbf{330}, 6003, p. 463-468 doi: 10.1126/science.1186986 (2010).


\bibitem{Bandfield}
J. L. Bandfield, M. J. Poston, R. L. Klima, and C. S. Edwards, Widespread distribution of OH/H2O on the lunar surface inferred from spectral data, \textit{Nat. Geosci.}, \textbf{11}, p. 173–177, doi: 10.1038/s41561-018-0065-0 (2018).

\bibitem{Li}
S. Lia, P. G. Lucey, R. E. Milliken, P O. Hayne, E. Fisher, J.-P. Williams, D. M. Hurley, and R. C. Elphic, Direct evidence of surface exposed water ice in the lunar polar regions, \textit{PNAS}, \textbf{115}, 36, p.8907-8912, doi: 10.1073/pnas.1802345115 (2018).

\bibitem{Qiao}
L. Qiao, Z. Ling, J. W. Head,  M. A. Ivanov, and B. Liu, Analyses of Lunar Orbiter Laser Altimeter 1,064‐nm Albedo in Permanently Shadowed Regions of Polar Crater Flat Floors: Implications for Surface Water Ice Occurrence and Future In Situ Exploration, \textit{Earth and Space Science}, \textbf{6}, 3, p. 467-488, doi: 10.1029/2019EA000567 (2019).

\bibitem{Rubanenko}
L. Rubanenko, J. Venkatraman, and D. A. Paige, Thick ice deposits in shallow simple craters on the Moon and Mercury, \textit{Nat. Geosci.}, \textbf{12} p. 597-601, doi: 10.1038/s41561-019-0405-8 (2019).

\bibitem{Nash}
E. Sefton-Nash, J.-P. Williams, B. T. Greenhagen, T. J. Warren, J. L. Bandfield, K. M. Aye, F. Leader, M. A. Siegler, P. O. Hayne, N. Bowles, and D. A. Paige, Evidence for ultra-cold traps and surface water ice in the lunar south polar crater Amundsen, \textit{Icarus}, \textbf{332}, p. 1-13, doi: 10.1016/j.icarus.2019.06.002 (2019).

\bibitem{Hayne}
P. O. Hayne, A. Hendrix, E. Sefton-Nash, M. A. Siegler, P. G. Lucey, K. D. Retherford, J.-P. Williams, B. T. Greenhagen and D. A. Paige, Evidence for exposed water ice in the Moon’s south polar regions from Lunar Reconnaissance Orbiter ultraviolet albedo and temperature measurements, \textit{Icarus}, \textbf{255}, p. 58-69, Doi: 10.1016/j.icarus.2015.03.032 (2015).

\bibitem{Gladstone}
G. R. Gladstone, and 19 co-authors, Far-ultraviolet reflectance properties of the Moon’s permanently shadowed regions, \textit{J. Geophys. Res.}, \textbf{117}, E00H04, doi: 10.1029/2011JE003913 (2012).

\bibitem{Hapke}
B. Hapke, E. Wells, J. Wagner, W. and Partlow, Far-UV, visible, and visible reflectance spectra of frosts of H2O, CO2, NH3 and SO2, \textit{Icarus}, \textbf{47}, p.361-367, doi: 10.1016/0019-1035(81)90184-6 (1981).

\bibitem{WB}
S. G. Warren and R. E. Brandt, Optical constants of ice from the ultraviolet to the microwave: a revised compilation, \textit{J Geophys Res}, \textbf{113}, D14220, doi:10.1029/2007JD009744 (2008).

\bibitem{Jake}
J. L. Kloos, J. E. Moores, P. J. Godin, and E. Cloutis, Illumination conditions within permanently shadowed regions at the lunar poles: implications for in-situ passive remote sensing, \textit{Companion paper submitted to Acta Astronautica} (2020). 

\bibitem{Jake2}
J. L. Kloos and J. E. Moores, Mapping the Limited Extent of Earthshine within Lunar PSRs, \textit{Research Notes of the AAS}, \textbf{3}, 9, doi: 10.3847/2515-5172/ab4195 (2019).

\bibitem{Glenar}
D. A. Glenar, T. J. Stubbs, E. W.Schwieterman, T. D. Robinson, and T. A. Livengood, Earthshine as an illumination source at the Moon, \textit{Icarus}, \textbf{321}, p.841-856, doi: 10.1016/j.icarus.2018.12.025 (2019).

\bibitem{Goguen}
 M. S. Gudipati and J. Castillo-Rogez, The Science of Solar System Ices, \textit{Springer New York}, ISBN 9781461430766 (2013).



\end{thebibliography}
\end{document}